\documentclass[12pt]{iopart}

\usepackage{iopams}  
\usepackage{graphicx}  
\usepackage{graphics}  
\usepackage{psfrag}
  
\newcommand{\be}{\begin{equation}}      
\newcommand{\ee}{\end{equation}}      
      
\newcommand{\bef}{\begin{figure}}      
\newcommand{\eef}{\end{figure}}      
\newcommand{\bea}{\begin{eqnarray}}    
\newcommand{\eea}{\end{eqnarray}}

\def\bk{{\bf k}}  
\def\bx{{\bf x}}  

\def\br{{\bf r}}

\def\bF{{\bf F}}

\def\bu{{\bf u}}

\def\lsim{\raise 0.4ex\hbox{$<$}\kern -0.8em\lower 0.62ex\hbox{$\sim$}} 
\def\gsim{\raise 0.4ex\hbox{$>$}\kern -0.7em\lower 0.62ex\hbox{$\sim$}}

\def\bi{\mathbf{i}}

\def\bk{\mathbf{k}}

\def\f0N{f_0^{(N)}}
\def\bec{\begin{center}}
\def\eec{\end{center}}

\newcommand{\ve}[1]{\ensuremath{\mathbf{#1}}}
\newcommand{\D}[1][ ]{\ensuremath{\mathrm{d}^{#1} }}

\begin{document}

\title[Dynamics of finite and infinite self-gravitating systems
]
{Dynamics of finite and infinite self-gravitating systems with  
cold quasi-uniform initial conditions}

\author[M. Joyce, B. Marcos and  F. Sylos Labini]
{M. Joyce$^1$, B. Marcos${^{2,}}{^{3,}}{^4}$ and F. Sylos Labini${^{2,}}{^{3}}$\\
}

\address{
$^1$Laboratoire de Physique Nucl\'eaire et de Hautes Energies,
UMR 7585, Universit\'e Pierre et Marie Curie --- Paris 6, 
75252 Paris Cedex 05, France\\
$^{2}$``E. Fermi'' Center, Via Panisperna 89 A, Compendio del
Viminale, I-00184 Rome, Italy\\
$^3$ISC-CNR, Via dei Taurini 19, I-00185 Rome, Italy\\
$^4$Laboratoire J.-A. Dieudonn\'e, UMR 6621, 
Universit\'e de Nice --- Sophia Antipolis,
Parc Valrose 06108 Nice Cedex 02, France}
\begin{abstract}
Purely self-gravitating systems of point particles have been
extensively studied in astrophysics and cosmology, mainly through
numerical simulations, but understanding of their dynamics still
remains extremely limited.  We describe here results of a detailed
study of a simple class of cold quasi-uniform initial conditions, for
both finite open systems and infinite systems. These examples
illustrate well the qualitative features of the quite different
dynamics observed in each case, and also clarify the relation between
them. In the finite case our study highlights the potential importance
of energy and mass ejection prior to virialization, a phenomenon which
has been previously overlooked. We discuss in both cases the validity
of a mean-field Vlasov-Poisson description of the dynamics observed,
and specifically the question of how particle number should be
extrapolated to test for it.
\end{abstract}

\maketitle

\section{Introduction}

Self-gravitating systems constituted by large numbers of classical 
point particles interacting by Newtonian gravity are still very 
poorly understood from a fundamental point of view. This is true
despite the extensive study of them in the context of astrophysics
and cosmology, where they are of central importance in realistic 
models of galaxy and large scale structure formation in the universe.
In cosmology, for example, the approximation of purely self-gravitating
particles is particularly important because, in currently favored models
for the universe, most of the self-gravitating matter is ``dark'', i.e.,
has extremely weak non-gravitational interactions, and further the
Newtonian approximation is valid in the regime in which structures
form. Indeed the canonical way in which predictions for the distribution
of galaxies in the universe are currently produced starts from a 
numerical simulation of particles in an infinite expanding space interacting
by purely Newtonian gravity (for a review, see e.g. \cite{dolag_etal_2008}). 
As we will discuss below, the non-expanding
limit of this problem (i.e. in a static infinite space) corresponds
to a particular regularization of the dynamics of such a finite 
self-gravitating system in the usual thermodynamic limit. The
case of an expanding (or contracting) infinite system, on the 
other hand, represents the limit in which the size of 
an expanding (or contracting) sphere tends to infinity.

While the ``real'' physical problem is thus the infinite
space one, the approximation in which astrophysical systems are
treated as finite is of course also very relevant: as long as the tidal
forces on a finite region coming from the rest of the universe 
may be neglected, it is a good approximation to treat the finite mass
in such a region in isolation. This corresponds to treating a finite
mass in an infinite space, i.e., with open boundary conditions. One 
can, of course, also study of the case of a finite 
self-gravitating system enclosed in a box. While such an
idealization may be interesting and instructive theoretically
(e.g. in allowing thermodynamic equilibrium to be defined \cite{padm}), 
its physical relevance is, however, less evident and we
will not consider it here.

We present here results of numerical simulations from simple controlled
sets of initial conditions, for  1) a finite
open self-gravitating system, and 2) an infinite self-gravitating
system. Our goal is to explain clearly the relation between these two
different cases, and to illustrate the typical phenomenology of the
dynamics in each case. We underline the open theoretical problems in
each case, and discuss some general questions about the framework in
which these problems should be addressed.  In particular our study shows 
that the assumption of energy and mass
conservation in violent relaxation of finite systems needs to be
considered with care, and we discuss also, for both cases, the question
of the validity of an appropriate mean-field Vlasov-Poisson system.  In both
cases we define an appropriate extrapolation which can be used to test
numerically for convergence to this limit.

The article is organized as follows. In the next section we describe
the phenomenology of the dynamical evolution of a finite open
self-gravitating system starting from a simple class of cold
(i.e. zero kinetic energy) initial conditions.  The qualitative nature
of the evolution is well known, and 
is generic in long-range interacting systems: through
``violent relaxation'' the system reaches a macroscopically stationary
state on a few dynamical time scales. We study the dependence of this
state on the initial conditions, which are uniquely parametrized, after
scaling, by the particle number $N$. In light of these behaviors we discuss the
features required of theoretical attempts to explain them, and in
particular the question of the validity of the Vlasov Poisson
limit. In Sect.~\ref{The infinite system limit} we then consider the
passage to the infinite system limit, and then in Sect.~\ref{Dynamics
  of an infinite system} we describe results of numerical simulations
of this limit for a simple class of initial conditions which
generalize appropriately those studied in the finite case. We describe
briefly the phenomenology of the completely out-of-equilibrium
dynamical evolution, highlighting that is shows the qualitative
features of ``realistic'', but more complicated, cosmological
models. In this case also we discuss the issue of the validity of the
Vlasov-Poisson limit for the dynamics of the discrete system, which is
in fact a question of practical importance in the exploitation of the
results of such simulations in cosmology.

\section{Dynamics of a finite open system}

We consider the evolution under their self-gravity of a finite
number $N$ of particles, i.e., with equations of motion
\be
\label{Nbody-eom}
\ddot\br_i=-{Gm}\sum_{j\ne i}\frac{\br_i-\br_j}{|\br_i-\br_j|^3}\,,
\,
\ee 
and open boundary conditions, where $\br_i$ is the position of the
$i$-th particle ($i=1..N$) and a dot denotes derivation with respect
to time. As initial conditions we 
take the $N$ particles {\it distributed randomly 
in a spherical region and at rest},  i.e., a sphere of cold matter 
with Poissonian density fluctuations.
The initial conditions are thus characterized by the single 
parameter $N$, or alternatively by the mean inter-particle 
separation $\ell$ defined as $\ell \equiv (3V/4\pi N)^{1/3}=R/N^{1/3}$,
where $V$ is the volume of the sphere of radius $R$.
As {\it unit of length} we will use the diameter of the 
initial sphere, i.e., $R=0.5$;
and as {\it unit of time } 
\be 
\label{tauscm}
\tau_{scm} \equiv \sqrt{\frac{3\pi}{32 G \rho_0}} \;,
\ee 
where $\rho_0$ is the initial mean mass density.


The significance of the time $\tau_{scm}$ is that it
corresponds to that at which a cold sphere of 
{\it exactly} uniform density  $\rho_0$
collapses to a density singularity. Indeed using Gauss' theorem
it is straightforward to show that such an initial 
condition gives an evolution described by a 
simple homologous contraction of the whole system, i.e., 
the position of any ``particle'' can be 
written as $\br (t)=a(t) \br(0)$ where $\br(0)$ is its
initial position (with respect to the center of the 
sphere). The ``scale factor'' $a(t)$ then obeys the
equation
\be
\label{frw-1}
\frac{\ddot{a}}{a}= -\frac{4\pi G}{3} 
\frac{\rho_0}{a^3} \,,
\ee
which can be integrated to give
\be
\left(\frac{\dot{a}}{a}\right)^2
=\frac{8\pi G \rho_0}{3}\left[\frac{1}{a^3}-\frac{1}{a^2}\right]\,, 
\label{hubble-cold}
\ee
of which the solution may be written in the parametric form
\bea
&& 
a(\xi) = \frac{1}{2} (1 + \cos \xi )
\\ \nonumber 
&&
t(\xi) 
= \frac{\tau_{scm}}{\pi} \left( \xi + \sin \xi \right)
\;.
\label{scm2}
\eea 
Thus all the mass collapses to a point as $R \rightarrow 0$
when $t \rightarrow \tau_{scm}$ ($\xi \rightarrow \pi$).
Note that this time is {\it independent of the size of the
system}. Equation (\ref{hubble-cold}) coincides, in fact, 
with that derived in general relativity for the scale 
factor of an {\it infinite} homogeneous collapsing 
universe containing only cold matter, and indeed this 
Newtonian solution for a finite homogeneous sphere 
is only a special case of a class of such solutions obeying 
the equation
\be
H^2(t) \equiv \left(\frac{\dot{a}}{a}\right)^2
=\frac{8\pi G \rho_0}{3}\left[\frac{1}{a^3}+\frac{\kappa}{a^2}\right] 
\label{friedmann}
\ee
where the dimensionless constant $\kappa$ is given by
\be
\kappa=  \frac{3H_0^2}{8\pi G \rho_0}  - 1 \,,
\ee
and $H_0=H(0)={\dot{a}(0)}/{a(0)}$ is a global 
``expansion rate''(describing a contraction if $H_0 <0$)
of the sphere at $t=0$. These solutions coincide 
with the full class of solutions for an infinite universe 
containing pressure-less matter in general relativity. 
The function $H(t)$ is then the Hubble ``constant'', defining 
the rate of global expansion (or contraction) of the universe. 
We will return to this relation between the Newtonian
problem for a finite system and the infinite space limit 
below.
 
While evolution from our chosen cold initial condition is well defined
at any finite $N$, the singularity at the finite time $\tau_{scm}$ in
the continuum limit, which corresponds simply to $N \rightarrow
\infty$ at constant mass density, makes it
expensive to numerically integrate these initial conditions as $N$
increases.  Indeed for this reason most numerical studies in the
astrophysical literature of spherical collapse have excluded it,
focusing instead on initial conditions with non-trivial inhomogeneous
distributions (e.g. radially dependent density) and/or significant
non-zero velocities (see \cite{mjbmfsl_halos2008} for references).  A
few studies of this case do, nevertheless, exist
\cite{aarseth_etal_1988, boily_etal_2002, morikawa_nongauss,
  morikawa_virial} (and will be discussed below), and that in
\cite{boily_etal_2002} gives results for $N$ as large as $10^7$.  Our
study covers a range of $N$ between several hundred and several
hundred thousand. Details of our numerical simulations performed using
the publicly available and widely used GADGET2 code
\cite{gadget,gadget_paper, gadget2005}, are given in \cite{mjbmfsl_halos2008}.
We mention here just one important consideration: instead of the exact
Newtonian potential, the code actually employs, for numerical reasons,
a two-body potential which is exactly Newtonian above a finite
``smoothing length'' $\varepsilon$, and regularized below this scale
to give a force which is 1) attractive everywhere and 2) vanishes at
zero separation. The (complicated) analytic expression for the
smoothing function may be found in \cite{gadget_paper}. With this
modified force the code does not integrate accurately trajectories in
which particles have close encounters, which lead to very large
accelerations and thus the necessity for very small time steps (which
is numerically costly). However, on the (short) time scales we 
consider such trajectories should not play any significant role
in modifying the {\it macroscopic} properties we are interested
in. The results shown below correspond to a constant value of
$\varepsilon$ in all simulations, a few times smaller than $\ell$ in
the largest $N$ simulation. As we will detail further below when we
discuss the Vlasov-Poisson limit for our system, we have tested our
results in particular for stability when $\varepsilon$ is extrapolated
to {\it smaller} values, and we interpret them to be indicative, on
the relevant time scales, of the $\varepsilon=0$, i.e., the exact
Newtonian, limit of the evolution given by Eq.~(\ref{Nbody-eom})
\footnote{As a test we have also performed simulations using
a code with a direct $N^2$ summation, and without any screening.
For the range of $N$ (up to a few thousand) for which we can run
this code over the same physical time-scale, we find excellent
agreement with the results obtained with GADGET2 with the
smoothing we have adopted (see \cite{mjbmfsl_halos2008} for
exact parameter values, as well as details of energy 
conservation etc.).}.

\subsection{Results}

\begin{figure}
\centerline{\includegraphics*[width=0.9\textwidth]{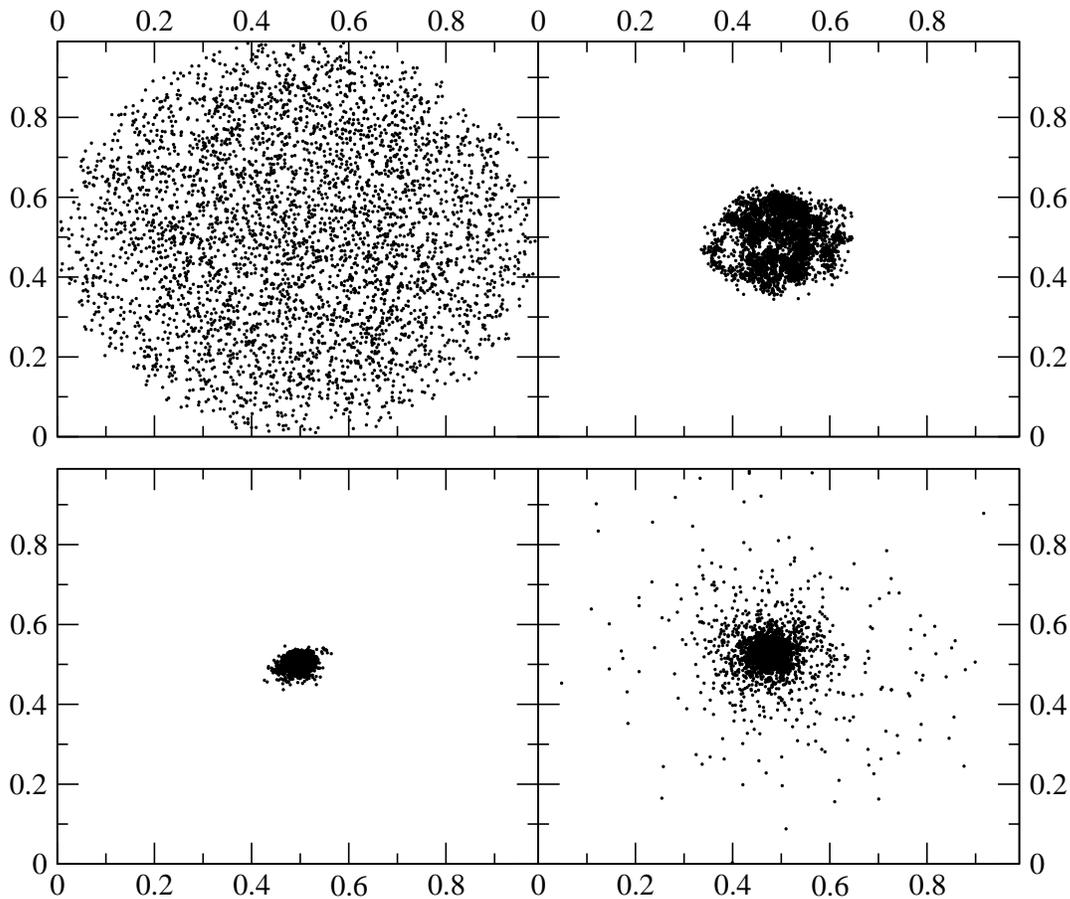}}
\caption{Snapshots of the evolution at from the cold sphere initial
  conditions described in the text.  The first three snapshots are 
  from the initial phase of contraction. Just after the third one
  (corresponding to $t=1$) the system reaches a minimal size, after
  which it  ``turns around'', ejecting some mass to infinity 
 and binding the rest in a virialized structure shown in the last
 snapshot.
\label{fig:collapse-finite}}
\end{figure}

Qualitatively the evolution we observe in all our simulations is the
same, and like that well known in both astrophysics, and, more
generally, in statistical physics for systems with long-range
attractive interactions from sub-virial initial conditions of this
type (i.e. with an initial virial ratio larger than -1) : the
system first contracts and relaxes ``violently'' (i.e. on timescales
of order the dynamical time scale $\tau_{scm}$) to give a virialized,
macroscopically stationary, state (see \cite{mjbmfsl_halos2008} for 
astrophysical references, and e.g. \cite{Dauxoisetal} for statistical
physics references). Such states are known variously as
``quasi-equilbria'', or ``meta-equilibria'' or ``quasi-stationary''
states, because they are understood not to be true equilibria,
but rather stable only on a time scale which diverges as some 
power of $N$ --- in proportion 
to $N/\log N$ for the case of gravity, according to simple
arguments \cite{binney}. Indeed on these much longer timescales ---
which we do not explore here --- the system is believed to
be intrinsically unstable to so-called ``gravothermal catastrophe''
\cite{lyndenbell+wood_1968} (see, e.g., \cite{klinko+miller_2004} 
for a recent exploration of this regime, and further references).
We will use here the term ``quasi-stationary states'', shortened to
QSS, in line with the predominant usage in the
statistical physics literature in the last few years.

Shown in Fig.~\ref{fig:collapse-finite} are four snapshots
(with particle positions projected onto a plane) of
a simulation with $N=4096$ particles at times
$t=0$, $t=0.9$, $t=1$ and $t=2.5$. While at 
$t=1$ the system is still in the phase of 
contraction, at the last time it has already
settled into the QSS. In Fig.~\ref{fig_virial}
we show, for the simulations indicated (the
notation $PN$ means $N$ Poisson distributed
particles), the evolution of the virial ratio
\be
\label{virial}
b(t) = \frac{2K^n}{W^n}
\ee
where $K^n$ is the kinetic energy of the particles with
negative energy, and $W^n$ the potential energy associated
with the same particles. As we will discuss below, 
particles with positive energy are ejected completely from the 
system and thus they are not considered in Eq.\ref{virial}. 
We observe that the relaxation is remarkably rapid
in time, with the bound particles ``settling down'' towards the QSS in
considerably less than a further dynamical time $\tau_{scm}$ after the
maximal ``crunch''.  \bef
\centerline{\includegraphics*[width=12cm]{FIG2.eps}}
\caption{The virial ratio [see Eq.(\ref{virial})] as a function of
  time for different simulations. In the insert panel is shown a
``zoom'' on the behavior of one of these curves around the time
of collapse.
\label{fig_virial}
}
\eef
A typical evolution of the radial density profile is shown in 
Fig.~\ref{fig_dpt}: until close to the maximal collapsed 
configuration, it maintains, approximately, the top-hat 
form of the original configuration and then in a very
short time changes and stabilizes to its asymptotic form
\footnote{The profile is calculated from the the center of 
mass of the particles with negative energy at any given 
time.}. We will discuss the latter form and its dependence on $N$
below.
\bef
\centerline{\includegraphics*[width=12cm]{FIG3.eps}}
\caption{Density profile of particles with 
negative 
total energy at different times during the collapse, for a simulation
with $N=65536$.
\label{fig_dpt}
}
\eef

The existing studies in the astrophysical literature of this class
of initial conditions \cite{aarseth_etal_1988, boily_etal_2002}
focus on how the singular collapse of the uniform spherical
collapse model is regulated at finite $N$, and in particular on
the scaling with $N$ in numerical simulations of the minimal 
size reached by the system. Indeed in the study 
of  \cite{aarseth_etal_1988} the ``points'' represent 
masses with extension (e.g. proto-stars) and the central question 
the authors wish to address is whether these masses survive
or not the collapse of a cloud of which they are the constituents.  
(We note also the  more recent study  
\cite{morikawa_nongauss, morikawa_virial} which focuses
on the velocity distributions of the QSS.)
This minimal radius $R_{min}$ may be defined in different ways,
e.g., as the minimal value reached by the radius, measured from
the center of mass, enclosing $90\%$ of the mass. Alternatively
it can be estimated as the radius inferred from the potential
energy of the particles, the minimal radius corresponding to
the maximal negative potential energy.
The behavior of $R_{min}$, 
determined by the first method, as a function of $N$ is 
shown in Fig.~\ref{fig_rmin}. The fitted line 
$R_{min} \propto N^{-1/3}$ is a behavior which has
also been verified in both \cite{aarseth_etal_1988} and 
\cite{boily_etal_2002}, the latter for an $N$ as large
as $10^7$. As explained in these articles it is a behavior 
which can be predicted from very simple considerations.
First, neglecting boundary effects (i.e. treating the limit of
an infinite sphere), it can be shown that small density 
perturbations, in the fluid approximation, grow in 
proportion to $a^{-3/2}$ for $a <<1$. Second, the 
initial amplitude of the relative density perturbations scale 
in proportion to $N^{-1/2}$. If we assume that the
uniform spherical collapse model breaks down when these 
perturbations to uniformity become of order unity ---
we arrive at the prediction $R_{min} \propto N^{-1/3}$.
The same estimate can be given more physically as 
requiring a balance between the pressure forces, associated
with the growing velocity dispersion, and the gravitational
forces. Note that, in fact, $R_{min} \approx \ell$ 
as in our units $\ell=0.5/N^{1/3}$, i.e. the minimal
size reached by the system is not just 
proportional to, but in fact approximately equal to,
the initial inter-particle separation.

\bef
\centerline{\includegraphics*[width=12cm]{FIG4.eps}}
\caption{ Behavior of the minimal radius $R_{min}$ attained,
determined as described in text,  as a function of $N$.
The solid line corresponds to the prediction described in
the text.
\label{fig_rmin}
}
\eef
Let us consider now the scaling with $N$ of various other
fundamental quantities, a question which has not been considered in
previous works. Specifically we focus on the macroscopic
characterization of the QSS formed in all cases. 
Let us consider first the mass and energy: while all particles 
start with a negative energy, a finite fraction can in fact 
end up with a positive energy. Given that they 
move, from very shortly after the collapse, in the 
essentially time independent potential of the 
virialized (negative energy) particles, they escape from 
the system. While evidently the ejected mass is bounded
above by the initial mass, the ejected energy is, in
principle, unbounded above as the gravitational self
energy of the bounded final mass is unbounded below.

\bef
\centerline{\includegraphics*[width=12cm]{FIG5.eps}}
\caption{ The behavior of $f^p(t)$, the fraction of the particles
with positive energy, as a function of time for two
  different simulations. A dependence on
  the number of particles is manifest.
\label{fig_fpt}
}
\eef

\bef
\centerline{\includegraphics*[width=12cm]{FIG6.eps}}
\caption{Behavior of the fraction of ejected particles as a function
of the total number of particles in the system. The solid line
is the phenomenological fit given by Eq.~(\ref{fposlog}).
\label{fig_fpos}
}
\eef

\bef
\centerline{\includegraphics*[width=12cm]{FIG7.eps}}
\caption{Observed behavior of the ratio $K^p/f^p$ as
a function of $N$. 
\label{kposfpos}
}
\eef
That ejection of mass and energy indeed takes place, that it is
non-negligible, and $N$ dependent, is shown in Figs.~\ref{fig_fpt},
~\ref{fig_fpos} and ~\ref{kposfpos}. The first figure shows the fraction
$f^p$ of the particles with positive energy as a function of time in
two different simulations, while the second shows the asymptotic value
of $f^p$ in each simulation as a function of $N$ (i.e. the value
attained on the ``plateau'' in each simulation after a few dynamical
times, corresponding to particles which are definitively ejected on
these time scales.)\footnote{We note that while some previous
works (see \cite{mjbmfsl_halos2008} for references) have noted
the ejection of some small fraction of the mass in similar
cases, the significance of the energy ejection as $N$ increases,
and its $N$ dependence, has not previously been documented.
Theoretical studies of the ejection of mass from a pulsating 
spherical system --- which is qualitatively similar to that
described below for the ejection observed here--- can be found in 
\cite{david+theuns_1989, theuns+david_1990}.}

Although $f^p$ fluctuates in different realizations with
a given particle number, it shows a very slow, but 
systematic, increase as a function of $N$, varying
from approximately $15\%$ to almost $35\%$ over the
range of $N$ simulated. A reasonably good fit 
is given by  
\be
\label{fposlog}
f^p(N) \approx  a+ b \log(N)\;,
\ee
where $a=0.048$ and $b=0.022$. Alternatively it can be fit 
quite well (in the same range) by a power law 
$f^p \approx 0.1 N^{0.1}$. Note that these fits cannot, 
evidently, be extrapolated to arbitrarily 
large $N$ (as the mass ejected is bounded above), and 
thus our study does not actually definitively determine
the asymptotic large $N$ behavior of this quantity
despite the large particle numbers simulated.
As we will discuss briefly below, however, the 
mechanism we observe for this mass ejection leads 
us to expect that the value of $f^p$ should saturate
when $f^p \sim 0.5$.

Fig. ~\ref{kposfpos} shows the ratio $K^p/f^p$, i.e.,
the kinetic energy {\it per unit ejected mass}, as a 
function of $N$. It has a much clearer and well defined
growth as a function of $N$, well fit by 
$K^p/f^p \propto N^{1/3}$. Note that for the
largest values of $N$ simulated $K^p$ is
{\it almost ten times} the initial (potential)
energy $E_0$ of the system. 

The increasing ejection of energy means that the
particles which remain in the QSS become more
and more bound as $N$ increases. Indeed using 
total energy conservation,
and the fact that both the final potential
energy of the ejected particles and that
associated with the interaction of the
bound and ejected particles is negligible,
we have
\be E_0 = W^n + K^p + K^n \;.
\ee 
Further, since the bound particles in the QSS are virialized
we have
\be
\label{energyapprox2}
2 K^n + W^n =0 \;.
\ee
Thus we have
\be
\label{energyapprox3}
W^n=-2K^n=2(E_0 - K^p) 
\ee 
so that we have the approximate
behavior $W^n \propto -N^{-1/3}$ (when
we neglect the slow observed variation
with $N$ of $f^p$). 

The dependence of the ejected energy on $N$ thus implies that the
macroscopic properties of the QSS also depend on $N$. Studying the
radial density profiles of the (approximately spherically symmetric)
QSS we find that they can always be fit well by the simple functional
form 
\be
\label{dpscm1}
n(r) = \frac{n_0}{\left(1+\left(\frac{r}{r_0}\right)^4\right)} \;.
\ee
The $N$ dependence is encoded in that of the two parameters $n_0$ and
$r_0$, which we find are well fit by $r_0 \propto N^{-1/3}$ and $n_0
\propto N^{2}$.  In Fig.~\ref{fig_dpN} we show the density profiles
for various simulations with different $N$ where the axes have been
rescaled using these behaviors.
\bef
\centerline{\includegraphics*[width=12cm]{FIG8.eps}}
\caption{Density profile of the virialized structure at a time $t
  \approx 4 \tau_{scm}$ for simulations with different number of
  particles.  The y-axis has been normalized by $N^2$ and the x-axis
  by $N^{-1/3}$ (see text for explanations).  The behavior of
  Eq.~(\ref{dpscm1}) is shown for comparison.
\label{fig_dpN}
}
\eef
It is simple to show that these scalings with $N$ of $n_0$ and $r_0$
are simply those which follow from those just given for $f^p$ and
$W^n$: using the ansatz Eq.~(\ref{dpscm1}) one has that the number
$N^n$ of bound particles is proportional to $n_0 r_0^3$ while the
potential energy $W^n$ is proportional to $m^2n_0^2 r_0^5$ where $m$
is the mass of a particle.  The best fit behaviors for $r_0$ and
$n_0$ thus correspond, since $m\propto 1/N$, to $N^n \sim N$ (i.e. a
constant bound mass, and therefore a constant ejected fraction of the
mass $f^p$) and $W^n \sim N^{1/3}$. More detailed fits to $n_0$ and
$r_0$ show consistency also with the very slow variation of $f^p$
observed.  In summary the $N$ dependence of the QSS manifests itself
to a very good approximation simply in a scaling of its characteristic
size in proportion to $R_{min}$, the minimal radius attained in the
collapse (which, as we have seen, is proportional to the initial
inter-particle separation).

\subsection{Dynamics of ejection}

\begin{figure}
\centerline{
\psfrag{X}[c]{$t$}
\psfrag{Y}[c]{Arbitrary units}
\includegraphics*[width=0.75\textwidth]{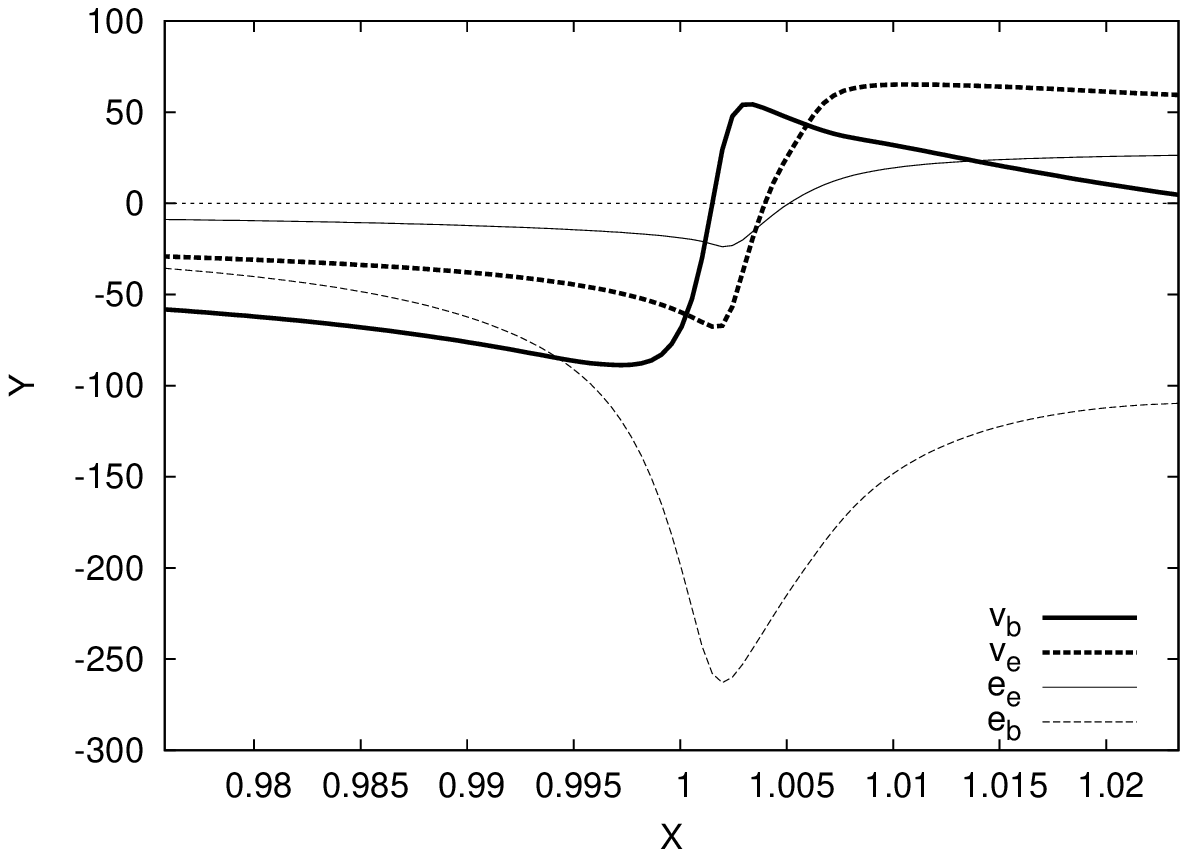}
}
\caption{Radial velocity, and average energy per particle, as a function
of time, of particles which are bound/ejected  at the end of the
simulation of $131072$ particles. The energy of the particles 
has been arbitrarily rescaled.
 \label{radial-131072}}
\end{figure}

We have investigated in detail the evolution of the system during
collapse and have identified the mechanism which leads to the mass and
energy ejection we have just described. We limit ourselves here to a
very brief qualitative description of our results, of which the full 
details are reported in \cite{mjbmfsl_halos2008}.  The probability 
of ejection is closely
correlated with particles' initial radial positions, with essentially
particles initially in the outer shells being ejected. The reason why
these particles are ejected can be understood as follows. Firstly,
particles closer to the outer boundary systematically lag (in space
and time) with respect to their uniform spherical collapse
trajectories more than those closer to the center.  This is an effect
which arises from the fact that, when mass moves around due to
fluctuations about uniformity, there is in a radial shell at the
boundary no average inward flux of mass to compensate the average
outward flux.  The mean mass density thus seen by a particle in such a
shell just decreases, leading to a slow of its fall towards the
origin. This ``lag'' with respect to particles in the inner shells
propagates in from the boundaries with time, leading to a coherent
relative lag of a significant fraction of the mass by the time of maximal
compression. Secondly, these lagging particles are then ejected as
they pick up energy, in a very short time around the collapse, as they
pass through the time-dependent potential of the particles initially
closer to the center, which have already collapsed and ``turned
around''.  This is illustrated in Fig.~\ref{radial-131072}, which
shows, for a simulation with $N=131072$ particles, the temporal
evolution of the components of the mass which are asymptotically
ejected or bound. More specifically the plot shows the evolution of
$v_e$ (and $v_b$) which is the average of the {\it radial} component
of the velocity for the ejected (and bound) particles, and also e$_e$
(and e$_b$) which is the mean energy per ejected (and bound) particle
(i.e. the average of the individual particle energies).  The
behaviors of $v_e$ and $v_b$ show clearly that the ejected particles
are those which arrive on average late at the center of mass, with
$v_e$ reaching its minimum after the bound particles have started
moving outward.  Considering the energies we see that it is in this
short time, in which the former particles pass through the latter,
that they pick up the additional energy which leads to their
ejection. Indeed the increase of $e_e$ sets in just after the change
in sign of $v_b$, i.e., when the bound component has (on average) just
``turned around'' and started moving outward again.  The mechanism of
the gain of energy leading to ejection is simply that the outer
particles, arriving later on average, move through the time dependent
{\it decreasing} mean field potential produced by the re-expanding
inner mass.  Assuming that the fraction of the lagging mass is
independent of $N$, an analysis of the scaling (see 
\cite{mjbmfsl_halos2008})
of the relevant
characteristic velocity/time/length scales allows one to infer the
observed scaling of the ejected energy with $N$. Quantitatively we
have not been able to explain, on the other hand, the observed 
$N$ dependence of the lagging mass, which should determine the $N$ dependence of
$f^p$. Given, however, that it is determined by a lag of the outer
mass relative to that of the inner mass, it seems clear that, as
required, the mechanism observed will naturally lead it to saturate at
a fixed fraction, of order one, as $N$ increases arbitrarily.

\subsection{Discussion}

In theoretical attempts to understand the properties of QSS produced
by violent relaxation using a statistical mechanics approach ---
whether the original one of Lynden-Bell \cite{lyndenbell} or variants
thereof --- two assumptions are generally made: 1) the mass and energy
of the virialized state is equal to the initial values of these
quantities, and 2) the dynamics is ``collisionless'', i.e., described
by the Vlasov-Poisson equations for the one particle phase space
density.

The first assumption is clearly not valid for the class of initial
conditions we have studied, not even as a first
approximation. Although the ejection we have observed, and in
particular its behavior with $N$, is clearly a result of the extreme
violence of the collapse due to the cold initial conditions, we do not
expect such ejection to be a feature only of this particular initial
condition. In a preliminary study of initial conditions incorporating
an initial velocity dispersion, we have found significant ejection
until an initial virial ratio of close to $0.5$ is reached. Thus, in
general, any such statistical approach to explaining the properties of
the virialized states should take into account in principle that the
dynamical evolution can change the values of the effective mass and
energy which is available for relaxation.  This observation is, we
believe, in line with the findings by Y. Levin et al. reported at this
conference (see also \cite{levin_etal_2008}): using studies of both
Coulomb and self-gravitating systems, they argue that the Lynden-Bell
approach works well when the violent relaxation is ``gentle'', but
that when it is violent the effects of resonances in the dynamics lead
to separation of part of the mass into a ``halo'' which does not relax
to the prescribed statistical equilibrium. We note that a similar
conclusion has been reached in earlier work on the $1-d$ ``sheet
model'' (see e.g. \cite{lecar+cohen}).  In the case we have studied
there is a ``dynamical resonance'' between the inner collapsing core
and the outer lagging particles in the collapse, leading to the
complete ejection of these particles from the system. In this context
we note that none of these findings should change if we consider, as
required notably in the Lynden-Bell framework, a system enclosed in a
finite volume instead of an open one: if we put our system in a box,
the ejected component will simply bounce off the walls and remain as
an unbound cloud moving in and out of the time independent potential
of the ``core''.

A second assumption of importance in theoretical models is that the
evolution of the system is well described by the coupled
Vlasov-Poisson (VP) equations (or ``collisionless Boltzmann equation''
coupled to the Poisson equation).  Is this the case? To determine
whether it is we need to understand how we can test for its
validity. As the VP limit is an appropriate $N \rightarrow \infty$
limit for the system, this means specifying precisely how this limit
should be taken. We can then extrapolate our numerical simulations to
larger $N$ to test for the stability of results.

It is clear that the appropriate extrapolation for the system we have
studied is not the naive limit $N \rightarrow \infty$, i.e., in which
we simply increase the number of Poisson distributed particles: we
have explicitly identified macroscopic $N$ dependencies in fundamental
quantities, so the results at any given $N$ do not approximate those
at any other $N$, and indeed do not converge towards any
$N$-independent behavior.

Formal proofs of the validity of the VP limit \cite{braun+hepp} for a
self-gravitating system require, however, that the singularity in the
gravitational force at zero separation be regulated when the limit $N
\rightarrow \infty$ is taken. This suggests we should take the limit
$N \rightarrow \infty$ while keeping fixed a smoothing scale, like the
$\varepsilon$ we have introduced in our simulations. Doing so we would
indeed expect to obtain a well defined $N$-independent result,
corresponding to the uniform spherical model with such a
regularization of the force: the sphere will not collapse below a
radius of order $\varepsilon$, as the force is then weaker than the
Newtonian force (and goes asymptotically to zero).  One would then
expect to obtain, for sufficiently large $N$, a final state which is
well defined and $N$ independent, but dependent on the scale of the
regularization and indeed even on the details of its implementation.

This limit is not the VP limit relevant here.  We have indeed
introduced a regularization of the force, but this has been done, as
we have discussed, only for reasons of numerical convenience, and our
criterion for our choice of $\varepsilon$ is that it be sufficiently
small so that our numerical results are independent of it.  Our
results are thus, a priori, independent of the scale $\varepsilon$
(and of how the associated regularization is implemented).  To
illustrate this we show in Fig.~\ref{fig_diffeps} the evolution of
$f^p$ (the fraction of particles with positive energy) as a function
of time, in simulations from identical initial conditions with
$N=32768$ particles in which only the value of $\varepsilon$ has been
varied, through the values indicated.  \bef
\centerline{\includegraphics*[width=12cm]{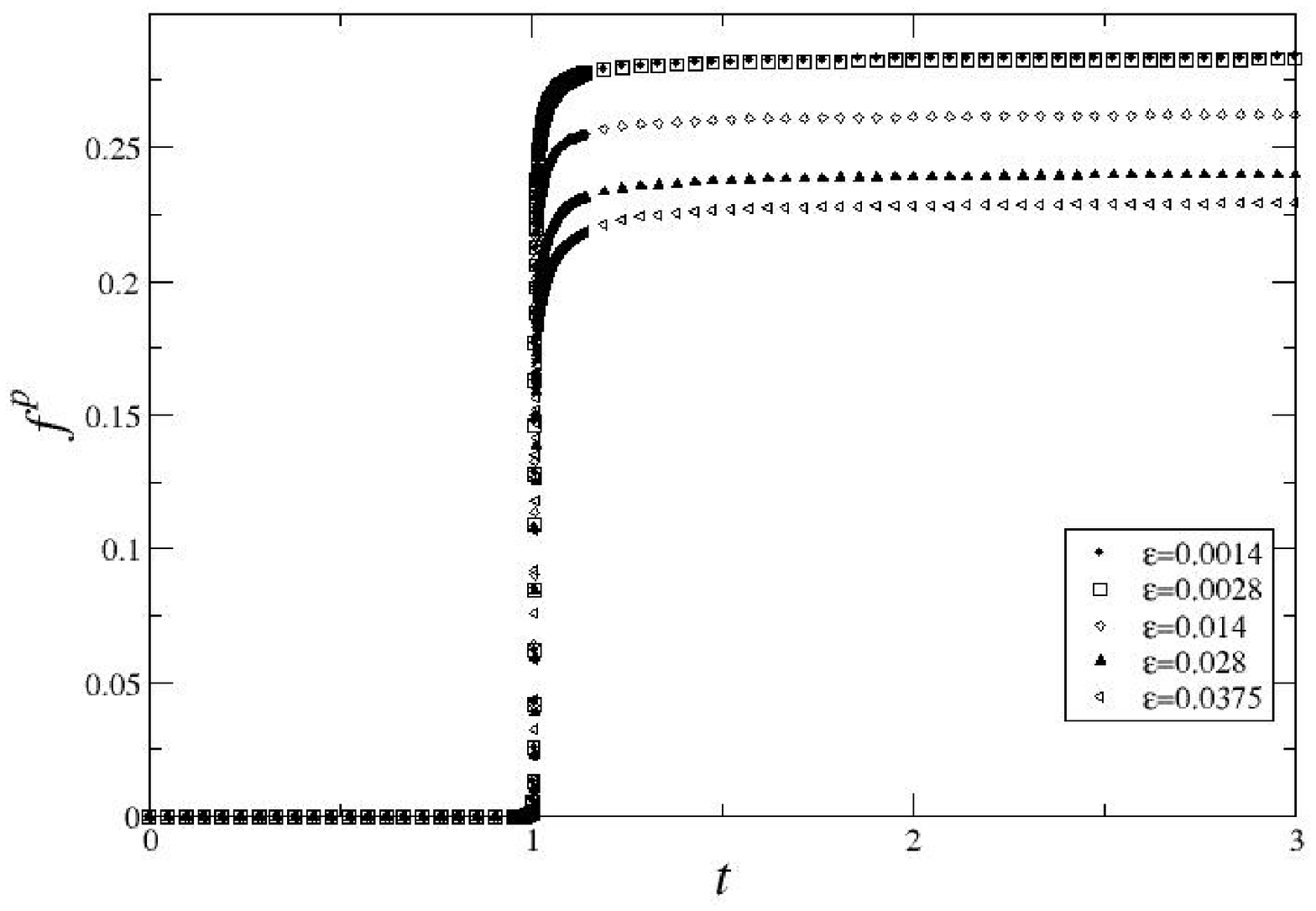}}
\caption{Evolution of the fraction of the mass with positive energy for
simulations with $N=32768$ for the different values indicated
of the smoothing parameter $\varepsilon$.
\label{fig_diffeps}
}
\eef
Other quantities we have considered show equally good convergence
as $\varepsilon$ decreases. Note that for the given simulation
the mean inter-particle distance in our units is $\ell=0.016$, so 
that the convergence of results is attained once
$\varepsilon$ is significantly less than $\ell$.  As we have 
seen, the minimal size reached by the collapsing system scales 
in proportion to $\ell$.  We interpret the observed convergence as 
due to the fact that the evolution of the system is determined primarily
by fluctuations on length scales between this scale
and the size of the system. Once $\varepsilon$ is sufficiently 
small to resolve these length scales at all times, convergence
is obtained.

Given the essential role played by fluctuations to the mean density in
determining the final state, it is clear that only an extrapolation of
$N$ which keeps the fluctuations in the initial conditions fixed can
be expected to leave the macroscopic results invariant. Any change in
$N$ necessarily leads, however, to some change in the fluctuations.
If, however, as indicated by the above results, there is a minimal
length scale in the initial conditions for ``relevant'' fluctuations,
we expect an extrapolation of $N$ which leaves fluctuations above this
minimal scale unchanged to give stable results.

Such an extrapolation for our initial conditions can be defined as
follows. Starting from a given Poissonian initial condition of $N$
particles in a sphere, we create a configuration with $N'=nN$
particles by splitting each particle into $n$ particles in a 
cube of side $2r_{s}$, centered on the original particle. The 
latter particles are distributed randomly 
in the cube, with the additional constraint that
their center of mass is located at the center of the cube, i.e.,
the position of the center of mass is conserved by the ``splitting''. 
In this new point distribution, which has the same mean
{\it mass} density as the original distribution,  fluctuations 
on scales larger than $r_s$ are essentially unchanged compared 
to those in the original distribution, while fluctuations around
and below this scale are modified (see
\cite{gabrielli+joyce_2008} for a detailed study of how fluctuations
are modified by such ``cloud processes''.). We have performed this
experiment for a Poisson initial condition with $N=4096$ particles,
splitting each particle into eight ($n=8$) to obtain an initial
condition with $N'=32768$ particles.  Results are shown in
Fig.~\ref{fig_split} for the ejected mass as a function of time, for a
range of values of the parameter $r_s$, expressed in terms of $\ell$,
the mean inter-particle separation (in the original distribution).
While for $r_s=0.8\ell$ the curve of ejected particles is actually
indistinguishable in the figure from the one for the original
distribution, differences can be seen for the other values, greater
discrepancy becoming evident as $r_s$ increases. This behavior is
clearly consistent with the conjecture that the macroscopic evolution
of the system depends only on initial fluctuations above some scale,
and that this scale is of order the initial inter-particle separation
$\ell$. And, as anticipated, this translates into an $N$ independence
of the results when $N$ is extrapolated in this way for an $r_s$
smaller than this scale.

This prescription for the VP limit can be justified theoretically
using a derivation of this limit through a coarse-graining of the
exact one particle distribution function over a window in phase space
(see e.g. \cite{buchert_dominguez}).  The VP equations are obtained
for the coarse-grained phase space density when terms describing
perturbations in velocity and force below the scale of the
coarse-graining are neglected. A system is thus well described by this
continuum VP limit if the effects of fluctuations below some
sufficiently small scale play no role in the evolution.  The
definition of the limit thus requires explicitly the existence of such
a length scale, and the limit is approached in practice when the mean
inter-particle distance becomes much smaller than this scale. With the
kind of procedure given we have defined not only an extrapolation of
$N$ which gives stable results, but also a method of identifying this
scale.

\bef 
\psfrag{X}[c]{ \large$t$}
\psfrag{Y}[c]{ \large$f^p$}
\centerline{\includegraphics*[width=12cm]{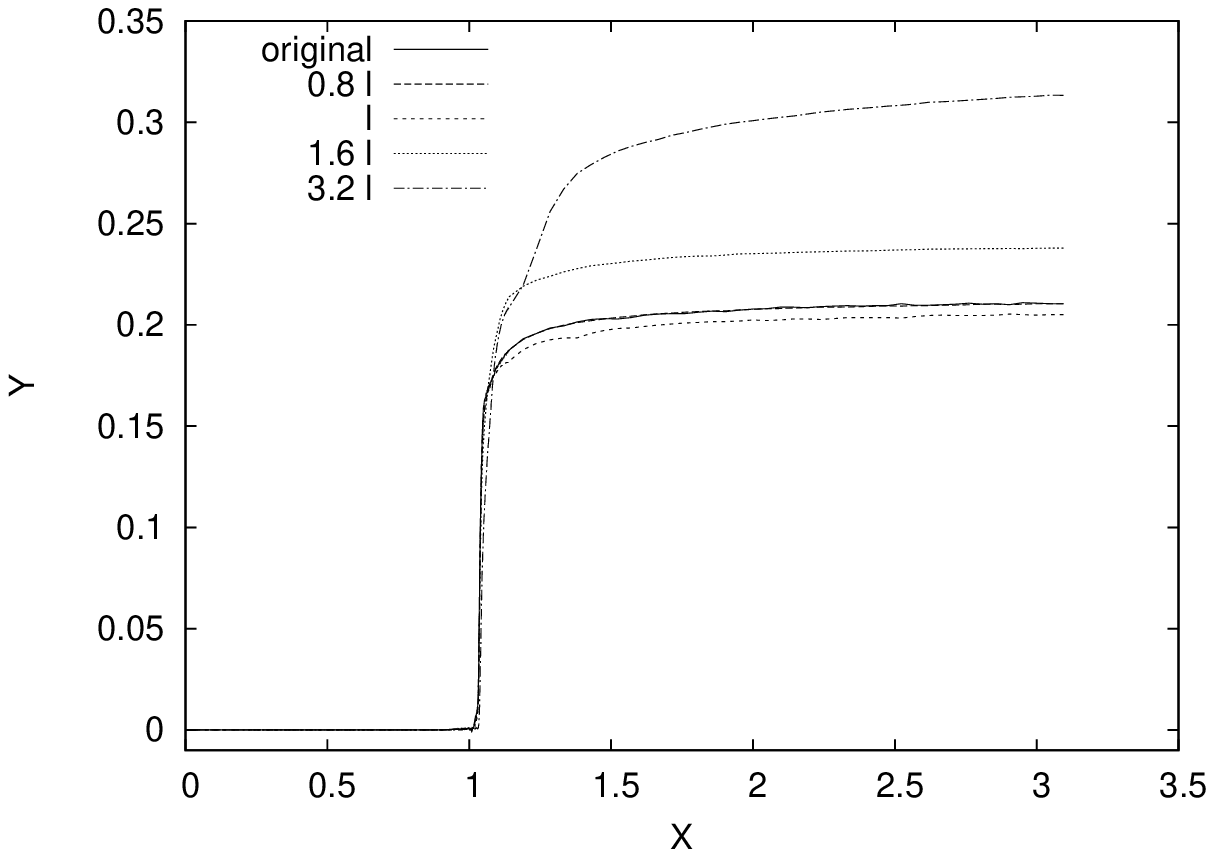}}
\caption{Evolution of the fraction of particles with positive energy 
as a function of time, for the different indicated values of the
parameter $r_s$ described in text. The ``original'' initial conditions
has $4096$ particles while the others have $32768$ particles.
The curve for $r_s=0.8$ is not visible because it is superimposed
on that of the original distribution.
\label{fig_split}}
\eef

\section{The infinite system limit}
\label{The infinite system limit}

Let us consider now the infinite system limit, i.e. the usual
thermodynamic limit in which $N \rightarrow \infty$ and
$V \rightarrow \infty$ at fixed particle density. This is
the limit relevant in the context of structure formation 
in the universe. In a Newtonian framework this limit is not
uniquely defined for the simple reason that the
sum giving the force on a particle on the right hand
side of Eq.~(\ref{Nbody-eom}) is not defined when
the distribution of points summed over is infinite 
with a non-zero mean density. By giving appropriate 
prescriptions for the calculation of this sum  --- or
equivalently, by giving a prescription for how the size
of the system is extrapolated --- one can define, however,
the dynamical problem in an infinite distribution. There
are two simple possibilities for such a prescription. One
gives exactly the equations for particle motion used in
cosmological $N$- body simulations of the expanding 
universe, the other defines the same problem in
a static universe. Although the latter of is less direct
practical relevance, it is, as we will discuss, an interesting
case for theoretical study, providing a simplified framework
in which to approach the full cosmological problem.
 
The case of an expanding (or contracting) universe is
simply a limit of the finite problem we have just studied.
The uniform cold spherical collapse model, which is the 
simple $N \rightarrow \infty$ limit of the finite initial 
condition we have studied, is, as we have discussed, 
singular at $t=\tau_{scm}$. At all times $t < \tau_{scm}$
it is, however, well defined for any size of the (spherical)
system,  and independently of this size. 
Until a time of order the maximal collapse in any 
finite $N$ system the trajectories of the particles 
are, to a first approximation, those in the 
uniform limit and it is thus useful to study the
problem in  ``comoving coordinates'' defined,
for the $i$-th particle with position vector
$\br_i (t)$ by $\bx_i (t) =  \br_i (t)/a(t)$.
The equations of motion Eqs.~(\ref{Nbody-eom})
then become, using Eq.~(\ref{frw-1}), 
\be
\ddot\bx_i +\frac{\dot a}{a} \dot\bx_i = -\frac{1}{a^3} 
\left[ \sum_{j\ne i} 
\frac{Gm(\bx_i-\bx_j)}{|\bx_i-\bx_j|^3} -\frac {4\pi G}{3} \rho_0 \bx_i \right] \,.
\label{finite-eom-comoving} 
\ee 
Now, in contrast to the force in Eq.~(\ref{Nbody-eom}), the expression
inside the brackets on the right-hand side, can be shown to be well
defined in a broad class of infinite point distributions: the
subtracted term removes the divergent term in the force arising from
the average density $\rho_0$. Specifically, for example, this
regularized force is known to be well defined in an infinite Poisson
distribution, in which case it is characterized by the so-called
Holtzmark distribution first derived by Chandrasekhar
\cite{chandra43}.
 
We can thus define an infinite volume limit for the
system, as that in which the equations of motion 
are given by Eqs.~(\ref{finite-eom-comoving}) with
the sum in the ``force'' taken over an appropriate 
infinite system (i.e. in the class in which it is 
defined). In order that $a(t)$ have the physical
significance it has in the case of a finite sphere,
the prescription on the force is that it be calculated
by summing in spheres about the chosen origin. 
The generalization to an expanding universe is obtained 
simply by taking an  appropriate expanding initial 
condition on the motion of the particles, i.e., with initially
outward velocities proportional to the particles'
positions with respect to the origin. As we noted
above this gives the appropriate Friedmann equation
(\ref{friedmann}) for $a(t)$. The case $\kappa=0$
corresponds to the so called Einstein-de Sitter 
universe, with $a(t) \propto t^{2/3}$. 

With respect to the problem we have studied above, the
infinite space limit defined in this way thus describes 
the development of the clustering inside the 
contracting (or expanding) sphere in the temporal
regime in which the finite-size effects we have 
described briefly above --- which lead to the
regularization of the singularity in the finite
case --- are negligible. Indeed in our simulations 
we find that the finite size 
effects we identified, of systematic particle lag
with respect to the uniform spherical collapse model,
propagate {\it at a given time} a radial distance 
into the interior of sphere which decreases  
as $N$ increases. Thus as we take the boundary
of the sphere to infinity we approach a well defined
and non-trivial limit for the particle trajectories
at any time $t < \tau_{scm}$, in which the regime
we have studied for the finite system simply 
disappears.

When the regulated force term on the right-hand side 
of Eqs.~(\ref{finite-eom-comoving}) is well defined
in an infinite distribution, they can be conveniently
rewritten as 
\be
\ddot\bx_i +\frac{\dot a}{a} \dot\bx_i =  
 -\frac{Gm}{a^3} 
\lim_{R_s \rightarrow \infty}  \sum_{j\ne i, j \in V(R_s, \bx_i)} 
\frac{\bx_i-\bx_j}{|\bx_i-\bx_j|^3} \,, 
\label{Nbody-eom-symm}
\ee where $V(R_s, \bx)$ is a sphere of radius $R_s$ centered on the
point with comoving position $\bx$. When the sum is performed in this
way --- symmetrically about each particle --- the contribution from
the mean density vanishes. These are the equations used in simulations
of structure formation in the universe, with 
the particles initially perturbed slightly
from an infinite perfect lattice, with small velocities in comoving
coordinates. (The infinite system is treated, in practice, as we
will discuss below, using the ``replica method'' in which the force 
is calculated in an infinite periodic system).

While, following the derivation given of these equations, the function
$a(t)$ should be a solution of Eq.~(\ref{friedmann}), when the system
is treated directly in comoving coordinates this function is in
practice inserted ``by hand'' (for the appropriate cosmology) and the
results for the clustering from given initial conditions depend on
it. Formally, however, one can insert {\it any} function $a(t)$ and study
the evolution of the clustering. A particularly interesting case,
which we will describe below, is when $a(t)=1$, i.e., a static
universe. It is not a solution to Eq.~(\ref{friedmann}) simply because
no exactly uniform distribution of purely self-gravitating matter in a
finite sphere (or indeed any geometry) is static.

This choice corresponds to what is known as the ``Jeans' swindle''
\cite{jeans, binney}: To treat the dynamics of perturbations to an
infinite self-gravitating pressure-full fluid, Jeans ``swindled'' by
perturbing about the state of constant mass density and zero velocity,
which is not, in fact a solution of these fluid equations. Formally
the ``swindle'' involves removing the mean density from the Poisson
equation, so that the gravitational potential is sourced only by
fluctuations to this mean density.  From Eq.~(\ref{Nbody-eom-symm})
with $a(t)=1$ it is manifest that this is equivalent to the
prescription that the gravitational force in an infinite distribution
be calculated by summing symmetrically about each point.  As discussed
by Kiessling~\cite{kiessling}, the presentation of what Jeans did as a
``mathematical swindle'' is in fact misleading: formulated in this way
as a regularization of the force --- that which makes the force zero
in an infinite uniform distribution --- it is perfectly well defined
mathematically.  The mathematical inconsistency in Jeans' analysis
arises because it is done in terms of potentials which are always
badly defined in the infinite volume limit, whereas forces, which are
the physically relevant quantities, may remain well defined. Kiessling
notes that an equivalent form of the Jeans' ``swindle'' prescription
for the force on a particle is \be
\label{symmetric-force-defn-screening}
\bF (\br_i) = -{Gm} \lim_{\mu \rightarrow 0}
\sum_{j\ne i} \frac{\br_i-\br_j}{|\br_i-\br_j|^3} e^{-\mu|\br_i-\br_j|}
\ee
where the sum now extends over the infinite space. This prescription
is an alternative to the one given above, which employs a 
sharp spherical top-hat window.  Both prescriptions (which give the 
same result, identical to the Jeans' ``swindle'') are simply regularizations
of the Newtonian force in the infinite volume (thermodynamic) limit
different to the spherical summation prescription which 
we have seen lead to the expanding/contracting universe.
The formulation in terms of a limiting procedure on a screened
potential is, however, very useful in that it gives a clearer
physical meaning to this infinite system limit: it means
that the dynamics we observe in this limit --- which we
will describe for a simple class of initial conditions
below ---  is the dynamics one would observe in a
screened gravitational potential starting from such
an initial condition, {\it up to a time when the length 
scales which characterize this dynamics becomes of
order the screening length}. In practice, as we will now
see, the relevant scale is the ``scale of non-linearity''.

\section{Dynamics of an infinite system}
\label{Dynamics of an infinite system}

We now describe the dynamical evolution of an infinite, initially
quasi-uniform, distribution of self-gravitating particles in
a static universe, i.e., with equations of motion which
can be written as 
\be
\label{eom-force-defn-screening}
\ddot{\br}_i = -{Gm} \lim_{\mu \rightarrow 0} \sum_{j\ne i}
\frac{\br_i-\br_j}{|\br_i-\br_j|^3} e^{-\mu|\br_i-\br_j|}\,.  \ee In
practice one can simulate numerically, of course, the motion of only a
finite number of particles. Following the discussion in the previous
section, we could in principle recover the relevant dynamics up to any
time from simulations of an appropriate sufficiently large finite
system with open boundary conditions.  Indeed we could use the
simulations discussed above for cold spherical collapse from
Poissonian initial conditions to study the dynamics of an infinite
contracting universe from Poissonian initial conditions: this dynamics
will be well approximated, as we have discussed, for a time prior to
the global collapse which increases as $N$ does. The case of a static
universe which we wish to consider, on the other hand, would require
that the gravitational interaction be screened on a scale considerably
smaller than the system size. This would prevent the global collapse,
but the dynamics will only approximate that in the infinite volume
limit as long as the clustering which develops is at a scale
sufficiently smaller than the screening scale.  Instead of such a
procedure it is much simpler to study numerically a system which is
infinite, but periodic, i.e., a finite cubic box and an infinite
number of copies.  This method, standard in cosmological simulations,
as well as in simulations of Coulomb systems in statistical physics,
is known as the ``replica method''. As the sum for the force in
Eq.~(\ref{eom-force-defn-screening}) is convergent it can be evaluated
numerically with desired precision in this infinite distribution. This
is usually done using the so-called Ewald sum method (see
e.g. \cite{hernquist}). The results given here have been obtained, as
above, using the publicly available GADGET2 code
\cite{gadget,gadget_paper,gadget2005}, which allows also the treatment of
infinite periodic systems based on this method.

As initial conditions for our study, which is reported in much greater
details in \cite{sl1, sl2, sl3}, we consider, for reasons we will now
explain, the following class of ``shuffled lattice'' (SL)
distributions, defined as follows: particles initially on a perfect
lattice, of lattice spacing $\ell$ (= mean inter-particle spacing),
are displaced randomly (``shuffled'') about their lattice site {\it
  independently} of all the others.  A particle initially at the
lattice site $\ve R$ is then at $\ve x(\ve R) = \ve R + \ve u(\ve R)$,
where the random vectors $\ve u(\ve R)$ are specified by $p(\ve u)$,
the PDF for the displacement of a single particle. We use a simple
top-hat form for the latter, i.e., $p(\bu) = (2\Delta)^{-3}$ for $\bu
\in [-\Delta,\Delta]^3$, and $p(\bu)=0$ otherwise . Taking $\Delta\to
0\,$, at fixed $\ell$, one thus obtains a perfect lattice, while
taking $\Delta\to \infty$ at fixed $\ell$, gives a Poisson particle
distribution~\cite{book}.

Given that we will be studying behaviors of the periodic system which
are independent of the size of the cubic box (i.e. representative of
the truly infinite SL as just defined), and that the (unsmoothed) 
gravitational interaction itself defines no characteristic scale, our 
chosen class
of initial conditions is actually {\it characterized by a single
  parameter}, which we may choose to be the dimensionless ratio
$\delta \equiv \Delta/\ell$, which we will refer to as the
\emph{normalized shuffling parameter}. Note that, in contrast to the
finite case, a Poissonian initial condition thus defines only a single
initial condition as in the infinite system limit it is characterized
solely by the mean inter-particle spacing which can always be taken as
unit of length (in the absence of a system size which defines an
independent length scale, and assuming that any small-scale 
smoothing of the interaction introduced plays no significant 
role in the dynamics on the time scales considered). 

To characterize the correlation properties of the SL, and those of the
evolved configurations, a useful quantity is the power spectrum
(or structure factor). We recall that for a point (or continuous mass)
distribution in a cube of side $L$, with periodic boundary conditions,
it is defined as  (see e.g. \cite{book})
\begin{equation}
P(\ve k) = \frac{1}{L^3}\langle | \tilde\delta(\ve k) |^2 \rangle \,,
\label{eq:pktheo}
\end{equation}
where $\tilde\delta(\ve k)$ are the Fourier components of the 
mass density contrast, which, for particles of equal mass, 
is simply
%
%
\begin{equation}
\tilde\delta(\ve k) = 
\frac{1}{n_0} \sum_{i=1}^{N} \exp(-i\ve k\cdot {\ve x}_i) \,,
\end{equation}
for $\ve k =(2\pi/L) \ve n$, where $\ve n$ is
a vector of integers, $n_0$ is the mean 
number density and the sum is over the $N$
particles (with the $i$-th at ${\ve x}_i$).
Note that with this normalization, which is 
that used canonically in cosmology, the asymptotic behavior 
at large $\bk$ in any point process is 
$P( \ve k \rightarrow \infty )=\frac{1}{n_0}$.
It is straightforward  to calculate
analytically the power spectrum for a generic SL. Expanded
in Taylor series about $\bk=0$ gives the leading small-$k$ behavior
\begin{equation}
P(\ve k) \approx \frac{k^2 \Delta^2}{3n_0} =\frac{1}{3} k^2 \delta^2
 \ell^5  \,,   
\label{eq:pk}
\end{equation}
where $k=|\bk|$. Note that this small-$k$ behavior of the 
power spectrum of the SL therefore does not depend on the 
details of the chosen PDF for the displacements,
but only on its (finite) variance. 

This class of SL initial conditions has the interest also
that, while simplified, they resemble those standardly used 
in cosmological simulations. In this context initial conditions 
are prepared by applying {\it correlated} displacements to a lattice.
By doing so one can produce, to a good approximation, a desired
power spectrum at small wave-numbers (See \cite{discreteness1_mjbm}
for a detailed description and analysis). The amplitude
of the relative displacements at adjacent lattice sites is
then related to the amplitude of the initial (very small) density 
fluctuations at the associated physical scale. In the
SL this translates into the fact that, from
Eq.~(\ref{eq:pk}), the amplitude of the 
power spectrum at small $k$ is a simple (quadratic)
function of the normalized shuffling $\delta$.

A full analysis of simulations for a range of different 
$\delta$ can be found in \cite{sl1}. As we wish here principally
to illustrate the qualitative features of the evolution
for a value of $\delta$ typical of a cosmological simulation
we will present results just for a single case, $\delta=1$.
We will then discuss the effect of the variation of $\delta$ 
in the context of our discussion of the validity of the 
VP limit. 

The results we give now are for two $\delta=1$ simulations: SL64, with
$64^3$ particles, and SL128, with $128^3$ particles of the same mass. 
We choose as unit of length that in which the box size in one in 
SL64, and two in SL128, i.e., the lattice spacing (and average 
mass density) is fixed and the 
simulations thus differ only in the box size. These simulations can 
be used to check that the dynamics is
indeed independent of the box size, and the results below are for the
temporal regime in which this is the case (i.e. in which the
quantities considered are in good agreement in the two
simulations). As in our finite system simulations we use, for
numerical efficiency, a non-zero smoothing $\varepsilon$, of which the
value in the SL64 and SL128 runs reported is $\varepsilon = 0.00175$
which corresponds to $\varepsilon \approx 0.1 \ell$.  As in this case
we have tested that our results are 
stable to the use of {\it smaller} values of $\varepsilon$.  The particles 
are again assigned zero velocity at the initial time, $t=0$.

\subsection{Results}

\begin{figure}
\centerline{\includegraphics*[width=\textwidth]{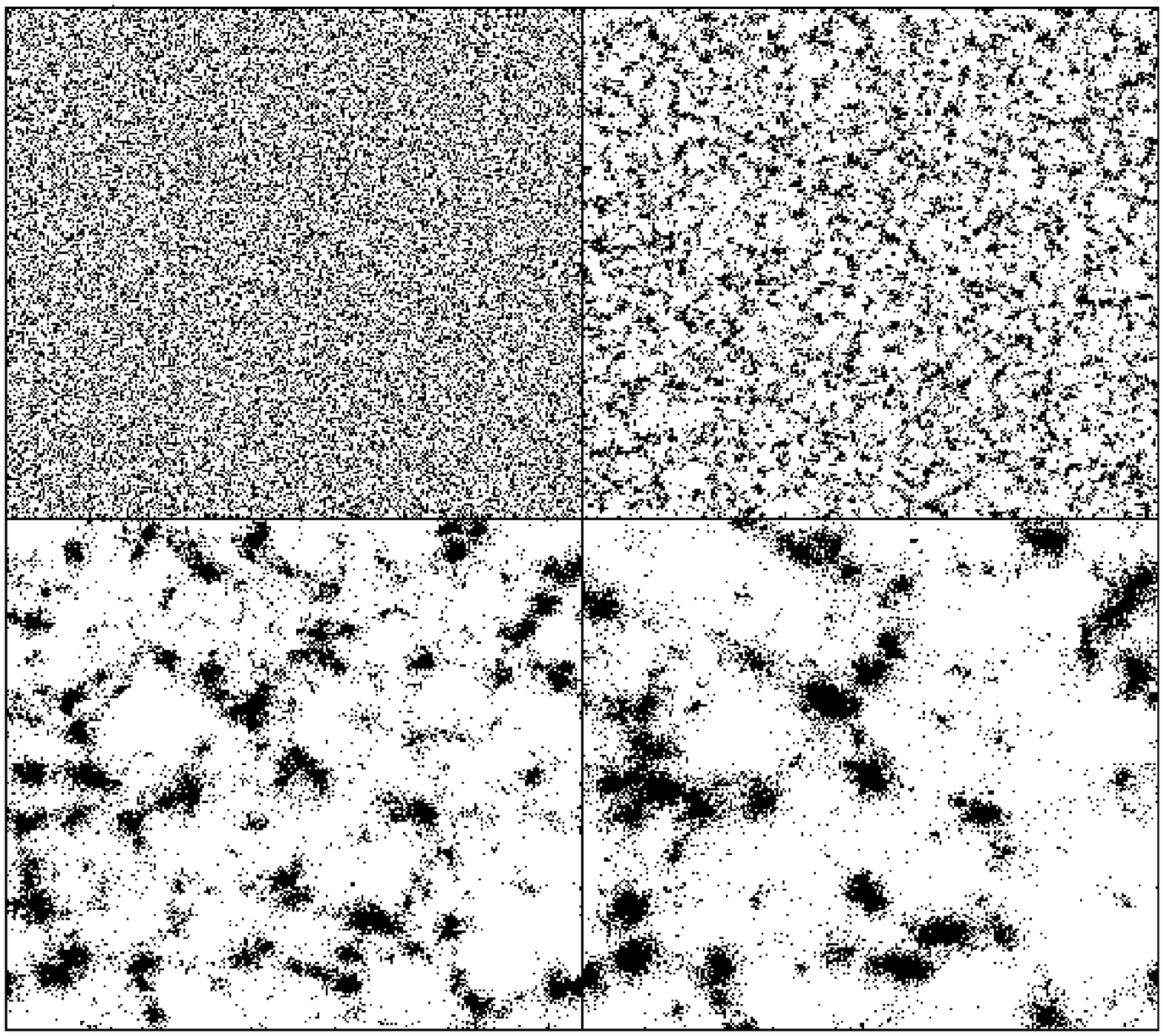}}
\caption{Snapshots of SL64, at $t=0$, and the evolved configurations
  obtained at subsequent times, t=$3,6,8$. (These are projections
onto the $x-y$ plane of a thin orthogonal slice of the full cube).
\label{fig:snapshots}}
\end{figure}

In Fig.~\ref{fig:snapshots} are shown four snapshots of the simulation
SL64. The time units here are slightly different than those used in
the finite spherical collapse simulations, the unit of time being
approximately $\tau_{scm}/2$ (see \cite{sl1} for the exact definition
employed).  We see visually that non-linear structures (i.e. regions
of strong clustering) appear to develop first at small scales, and
then propagate progressively to larger scales.  Eventually the size of
the structures become comparable to the box-size. From this time on
the evolution of the system will no longer be representative of that
in the infinite system.  Up to close to this time, however, it is
indeed the case that all the properties we will study below show
negligible dependence on the box size (see Refs.~\cite{sl1,sl2} for
more detail).

Let us consider first the evolution of clustering in 
real space, as characterized by the reduced correlation function $\xi(r)$. 
We recall that this is simply defined as 
\begin{equation} 
\xi(\ve r)= \langle \delta(\ve x+\ve r)\delta(\ve x)\rangle
 \;,
\end{equation}
where $\langle ... \rangle$ is an ensemble average, i.e., an average
over all possible realizations of the system (and we have assumed
statistical homogeneity). It is useful to note 
that this can be written, for $r\neq0$, and averaging over spherical
shells, 
\begin{equation}
\xi(r) = \frac{\langle n(r)\rangle_p }{n_0}-1 
\;,
\label{eq:xi_rdf}
\end{equation}  
where $\langle n(r)\rangle_p$ is the \emph{conditional average density},
i.e., the  (ensemble) average density of points in an infinitesimal
shell at distance $r$ from a point of the distribution. 
Thus $\xi(r)$ measures clustering by telling us how the density 
at a distance from a point is affected, in an average sense, by 
the fact that this point is occupied.  In distributions which 
are statistically homogeneous the power spectrum $P(\bk)$
and $\xi (\ve r)$ are a Fourier conjugate 
pair (see e.g. Ref.~\cite{book}).

The correlation functions here will invariably be 
monotonically decreasing functions of $r$. It is 
then useful to define the scale $\lambda$ by  
\begin{equation}
\xi(\lambda) = 1  \,.
\label{eq:homoscale}
\end{equation}
This scale then separates the regime of weak correlations
(i.e. $\xi(r) \ll  1$) from the regime of strong 
correlations (i.e. $\xi(r) \gg  1$). In the context of gravity 
these correspond, approximately, to what are referred
to as the {\it linear} and {\it non-linear} regimes,
as a linearized treatment of the evolution of
density fluctuations (see below) is valid in the former 
case.  Eq.~(\ref{eq:homoscale}) can also 
clearly be considered as a definition of 
the \emph{homogeneity scale} of the system. Physically
it gives then the {\it typical size of strongly clustered
regions}. 

\begin{figure}
\centerline{\includegraphics*[width=0.75\textwidth]{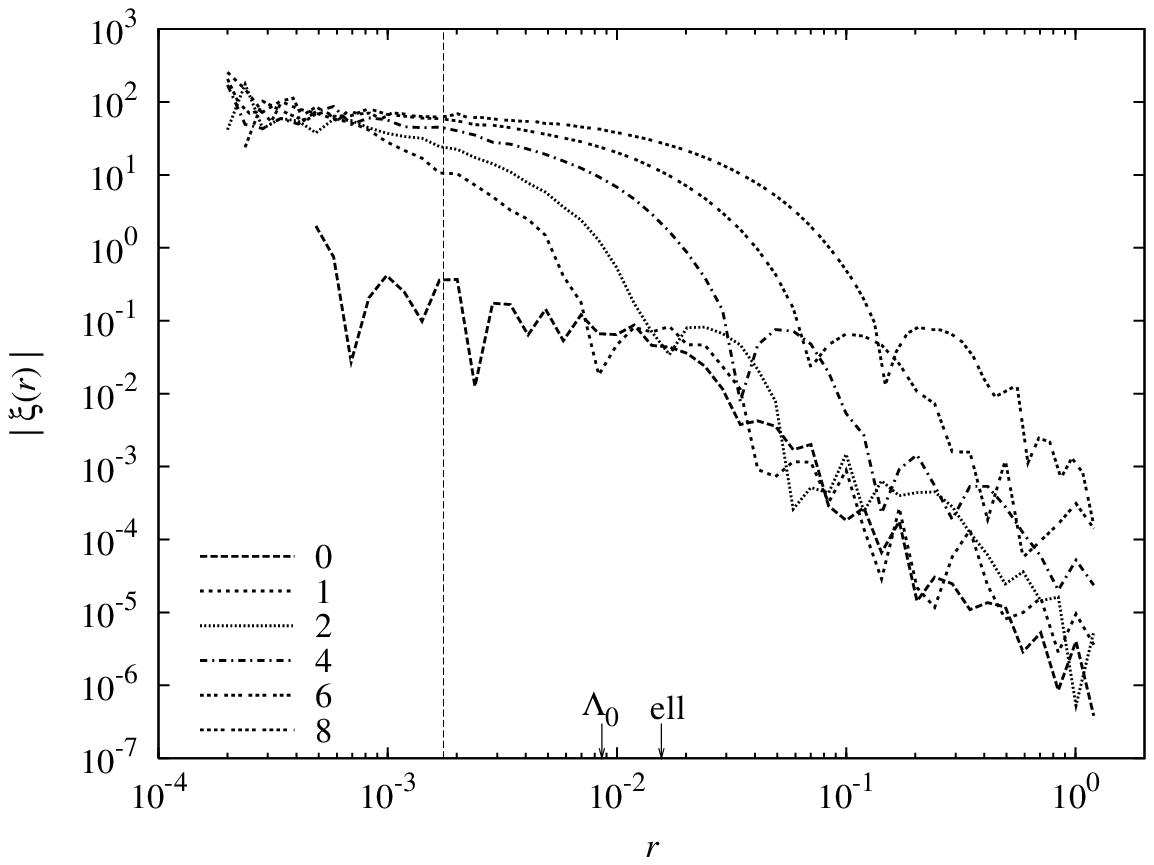}}
\caption{Behavior of the absolute
value of the correlation function $|\xi(r)|$ in SL128 at times
$t=0,1,2,4,6,8$. The vertical dotted line indicates $\varepsilon$. 
From \cite{sl1}.}
\label{fig:xi}
\end{figure}

In Fig.~\ref{fig:xi} is shown the evolution of the absolute value 
$|\xi(r)|$ in a log-log plot, for the SL128 simulation.
These results translate quantitatively the visual impression
gained above. More specifically we observe that:
\begin{itemize}
\item Starting from $\xi(r) \ll 1$ everywhere, non-linear correlations
(i.e. $\xi(r)\gg 1$ ) develop first at scales smaller than the
initial inter-particle distance.

\item After two dynamical times the clustering
develops little at scales below $\varepsilon$. The clustering around
and below this scale is characterized by an approximate ``plateau''.
This corresponds to the resolution limit imposed by the chosen
smoothing.

\item At scales larger than $\varepsilon$ the correlations
grow continuously in time at all scales, with the scale of
non-linearity [which can be defined, as discussed above, by
$\xi(\lambda)=1$] moving to larger scales.

\end{itemize}

Once significant non-linear
correlations are formed, the evolution of the correlation function
$\xi(r,t)$ can in fact be described, approximately, by a simple 
spatio-temporal scaling relation:
\begin{equation}
\xi(r,t) \approx \Xi\left( r/R_s(t) \right)\,, 
\label{eq:rescaling}
\end{equation}
where $R_s(t)$ is a time dependent length scale which we discuss in
what follows.
\begin{figure}
\centerline{\includegraphics*[width=0.75\textwidth]{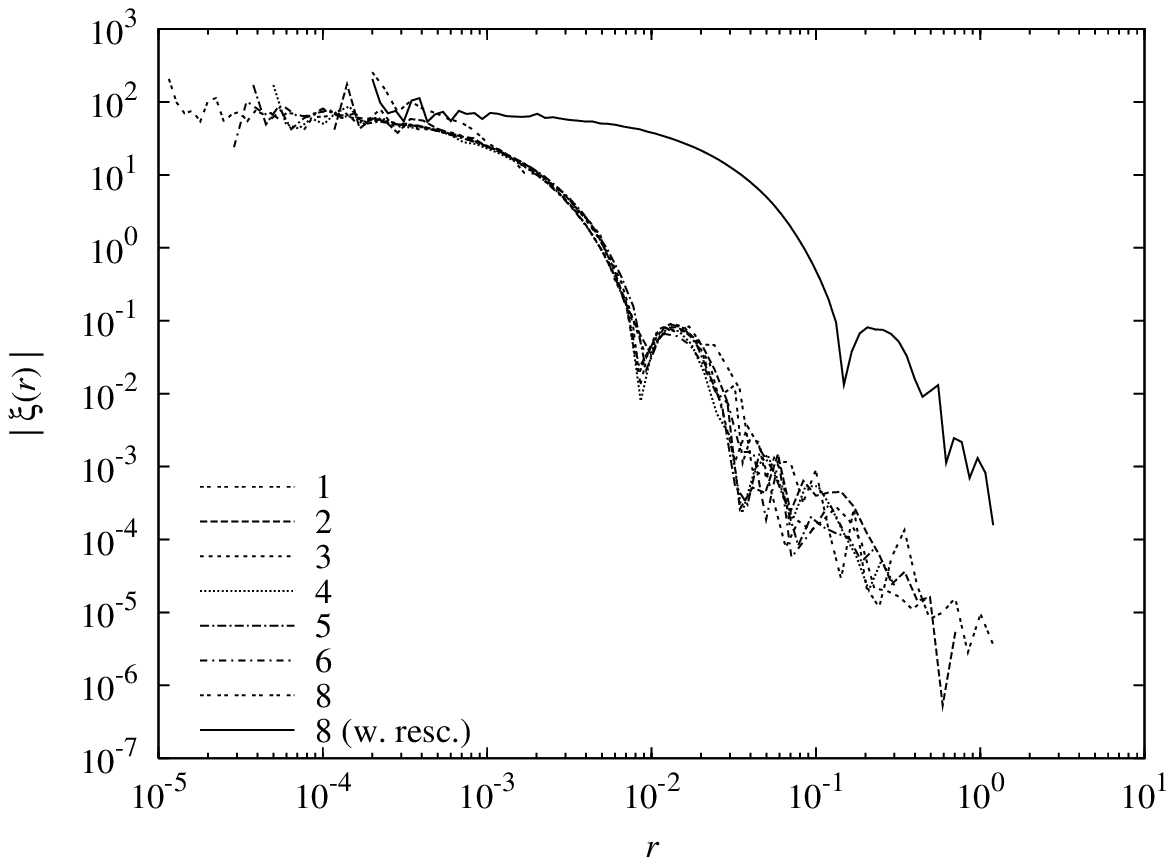}}
\caption{Collapse plot of $\xi(r,t)$: for each time $t>1$ we 
have rescaled the $x$-axis by a time-dependent factor to collapse all
the curves (dashed ones) to that at time $t=1$. We have added for
comparison $\xi(r,t=8)$ without rescaling (``w. resc.'', continuous
line). From \cite{sl1}.}
\label{fig:collapse}
\end{figure}
Shown in Fig.~\ref{fig:collapse} is ``collapse plot'' 
which allows one to evaluate the validity of this relation: $\xi(r,t)$
at different times is represented with a rescaling of the $x$-axis by
a (time-dependent) factor chosen to superimpose it as closely as
possible over itself at $t=1$, which is the time from which the
``translation'' appears to first become a good approximation.  We can
conclude clearly from Fig.~\ref{fig:collapse} that the relation
Eq.~(\ref{eq:rescaling}) indeed describes very well the evolution, down to
separations of order $\varepsilon$, and up to scales at which
the noise dominates the estimator (see \cite{sl1} for further details).
\begin{figure}
\centerline{\includegraphics*[width=0.7\textwidth]{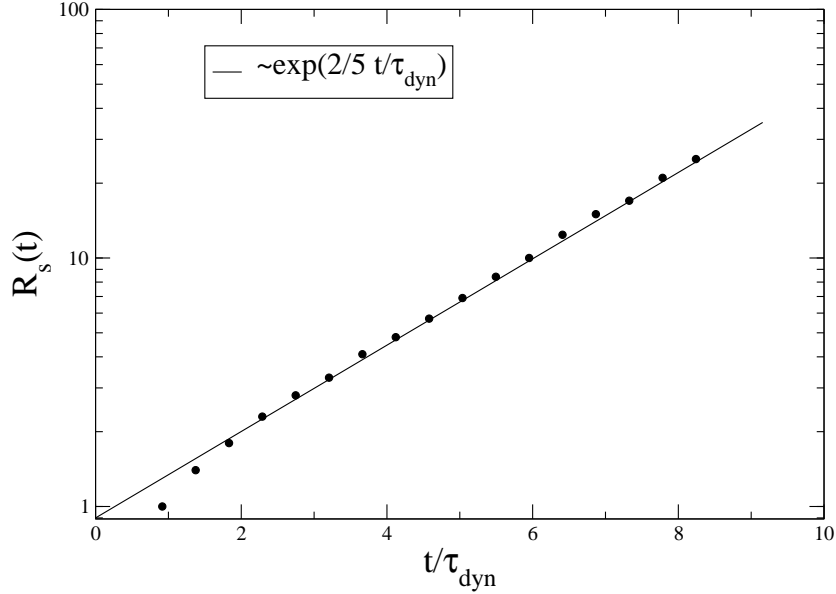}}
\caption{Evolution of the function $R_s(t)$ in SL128
(points) compared with its prediction from linearized
fluid treatment, as explained in the next section.
From \cite{sl1}.}
\label{fig:Rst}
\end{figure}

In Fig.~\ref{fig:Rst} is shown the evolution of the rescaling factor
$R_s(t)$ found in constructing Fig.~\ref{fig:collapse}, as a function
of time [with the choice $R_s(1)=1$]. 
The theoretical curve also shown, which fits 
the data very well at all but the earliest times, is that inferred from a 
simple analysis analogous to that used in cosmology 
for the expanding universe (see e.g. \cite{peebles}), 
which we now describe. In this context, this spatio-temporal
scaling behavior of the correlations is known 
as {\it self-similarity}. It has been observed in 
cosmological N body simulations starting from a range
of initial conditions characterized by a simple 
power-law PS (see e.g. \cite{efstathiou_88,bertschinger}).

\begin{figure}
\centerline{\includegraphics*[width=0.7\textwidth]{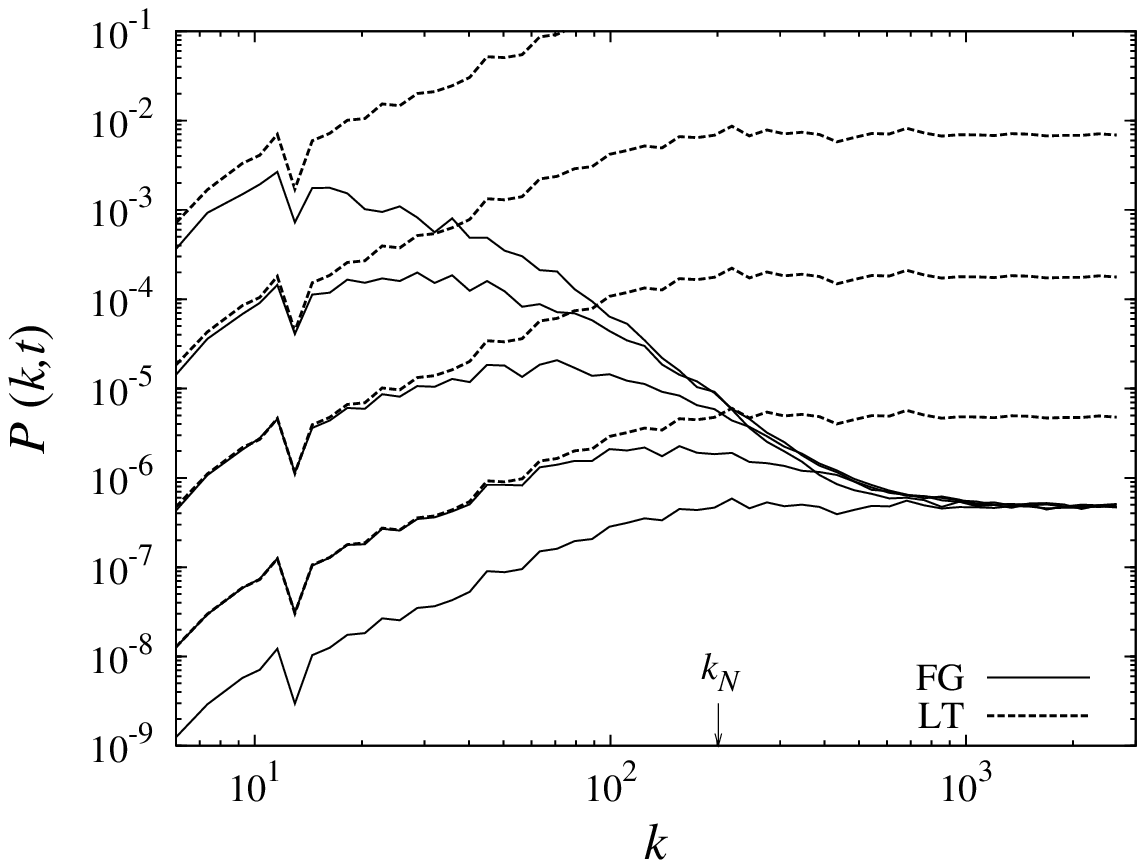}}
\caption{Evolution of the power spectrum 
in SL128 (solid lines --- label FG): the
  curves are for time equal to 0,2,4,6,8 (from bottom to up).  The
  dashed lines labeled with LT show the predictions of fluid linear
  theory, i.e., Eq.~(\ref{eq:pk_evolution_linear}) with $P(\bk,0)$
  measured in the simulation at $t=0$ for the same time steps. The
  arrow labeled ``$k_N$'' shows the value of the corresponding Nyquist
  frequency $k_N = \pi/\ell$. From \cite{sl1}.
\label{fig:pk_SL128} }
\end{figure}

Firstly we need the result of the standard ``linear theory''
in cosmology. When the equations for the 
density and velocity perturbations in an infinite self-gravitating 
pressure-less fluid (which can be obtained by an appropriate
truncation of the VP equations) are solved perturbatively, 
one obtains, in the static universe case, simply
\begin{equation}
  \ddot{\delta}(\ve x,t) = 4\pi G \rho_0 \delta(\ve x,t) \,\, {\rm and}\,\, 
  \ddot{\tilde\delta}(\ve k,t) = 4\pi G \rho_0 \tilde\delta(\ve k,t) \ 
\label{eq:ddotdelta}
\end{equation}
which, for the case of zero initial velocities, gives
\begin{equation}
  \delta(\bk,t) = \delta(\bk,0) \cosh\left( \sqrt{4\pi G \rho_0}\
  t\right) \;. \label{eq:densitycontrastlinear} 
\end{equation}
and thus for the power spectrum
\begin{equation}
P(\ve k,t) = P(\ve k,0) \cosh^2\left(t/\tau_{\rm {dyn}} \right) \,.
\label{eq:pk_evolution_linear}
\end{equation}
where we have defined $\tau_{\rm {dyn}}=1/\sqrt{4\pi G \rho_0}$.
That this indeed describes well the evolution of the power
spectrum at sufficiently small wave-numbers can be seen
from Fig.~\ref{fig:pk_SL128}, which shows this quantity
for the SL128 simulation along with the prediction of
Eq.~(\ref{eq:pk_evolution_linear}).

To obtain the prediction for $R_s(t)$ we now assume that
the required spatio-temporal scaling relation holds 
exactly, i.e., {\it at all scales},  from, say, a 
time $t_{s} > 0$. For $t > t_{s}$ we have then 
\bea
P(\ve k,t) &=&  \int_{L^3} \exp(-i\ve k\cdot \ve r) \,{\xi}(\ve
r,t) \, \D[3] \ve r \\
&=&  R_s^3(t) \int_{L^3} \exp(-i R_s(t) \ve k\cdot \ve x ) \,
\Xi(|\ve x|) \
\D[3] \ve x   \\
&=&  R_s^3(t) P(R_s(t)\ve k,t_s)  \ .
\label{eq:Rsdetermine}
\eea
where we have chosen $R_s(t_s)=1$. Assuming now that 
the power spectrum at small $k$ is amplified as given by linear theory,
i.e., as in Eq.~(\ref{eq:pk_evolution_linear}), one infers
for any power spectrum $P(k) \sim k^n$ that
\begin{equation} 
R_s(t) = 
\left( \frac{\cosh \frac{t}{\tau{\rm{dyn}}}}{\cosh \frac{t_s}{\tau{\rm{dyn}}}} \right)
^\frac{2}{3+n} 
\rightarrow
\exp\left[ \frac{2(t-t_s)}{(3+n)\tau{\rm{dyn}}} \right]
\,\, {\rm for} \,\, {t\gg t_s}\,.
\label{eq:predRst}
\end{equation}
In the asymptotic behavior the relative rescaling in space 
for any two times becomes a function only 
of the {\it difference} in time between them so that we can write
\begin{equation} 
\xi (r, t + \Delta t) = \xi \left(\frac{r}{R_s(\Delta t)}, t \right)
\; ; \quad R_s(\Delta t) = e^{ \frac{2 \Delta t} {(3+n) \tau{\rm{dyn}}} } \,.
\label{eq:self-sim-xi}
\end{equation}
The theoretical curve for $R_s(t)$ in  Fig.\ref{fig:Rst} 
corresponds to $n=2$ (for the SL initial condition) and
the best-fit choice of $t_s \approx 2.5$.

Let us make a few further observations about the evolution
of the power spectrum in Fig.~\ref{fig:pk_SL128}.
We observe that:
\begin{itemize}
\item 
The linear theory prediction describes the evolution very accurately
in a range $k < k^{*}(t)$, where $k^{*}(t)$ is a wave-number which
decreases as a function of time. This is precisely the qualitative
behavior one would anticipate (and also observed in cosmological
simulations): linear theory is expected to hold approximately
only above a scale in real space, and therefore up to some corresponding 
wave number in reciprocal scale, at which the averaged density 
fluctuations are sufficiently small so that the linear approximation 
may be made. A more precise study of the
validity of the linearized approximation is given in \cite{sl1}.
This scale in real space, as we have seen, clearly 
increases with time,  and thus in reciprocal space decreases with 
time. We note that at $t=6$ only the very smallest $k$-modes 
in the box are still in this {\it linear regime}, while at $t=8$ 
this is no longer true (and therefore finite size effects are expected
to begin to play an important role at this time).

\item
At very large wave-numbers, above $k_N$, the power spectrum remains 
equal to its
initial value $1/n_0$. This is simply a reflection of the necessary
presence of shot noise fluctuations at small scales due to the
particle nature of the distribution. 

\item 
In the intermediate range of $k$ the
evolution is quite different, and {\it slower}, than that given by
linear theory. This is the regime of {\it non-linear clustering}
as it manifests itself in reciprocal space.
\end{itemize} 

All these behaviors of the two point correlations are qualitatively
just like those observed in cosmological simulations in 
an expanding universe. More general than the ``self-similar''
properties (which apply only to power law initial conditions),
the clustering can be described as ``hierarchical'' , a feature
typical of all currently favored cosmological (``cold dark matter'' 
-type) models: structures develop at a scale which increases in
time, at a rate which can be determined from linear theory. This
is given the following physical interpretation: clustering may be 
understood essentially as produced by the collapse of small initial 
over-densities which evolve as prescribed by linear theory,
independently of pre-existing structures at smaller scales,
until they ``go non-linear''. 

\subsection{Discussion}

Given that the evolution of this simplified set of initial 
conditions in a static universe shows all the
qualitative features of that observed also in cosmological
$N$-body simulations, it provides an interesting ``toy model'' 
in which to address the many open problems in this context.
While fluid linear theory, which may be derived from a 
continuous Vlasov-Poisson description of the system, can 
account very well for the observed 
behavior of $R_s(t)$, the detailed nature of the 
non-linear regime of clustering remains very poorly
understood, with analytical approaches being essentially
limited to phenomenological models constructed from
simulations (e.g. ``halo models'', reviewed in \cite{halo}). 
The functional form of the two point correlation function 
attained in the asymptotic ``self-similar'' regime of the 
clustering is, for example, not understood. One fundamental 
question which is of relevance in this context is whether 
the relevant dynamics in the non-linear regime is also 
well described by the VP limit, i.e., by 
the Vlasov equations for the one particle phase space 
density with the acceleration calculated from
the ``Jeans' swindled'' Poisson equation (sourced only
by the density fluctuations). The question is also of
considerable practical importance in the context of
cosmological simulations for another reason: in this
context cosmological $N$-body simulations --- 
analogous to those we have 
described here --- are in fact employed simply because it is
not feasible numerically to simulate the VP
equations. The latter in fact represent the limit 
appropriate to describe the theoretical models, in
which the gravitating (``dark'') matter has a microscopic mass.
Thus the results of an $N$-body simulation are of
physical relevance only in so far as they do actually
represent well this limit. The problem of determining
the limitations on/accuracy of the $N$-body method 
is the ``problem of discreteness'' in cosmological  
$N$-body simulation, and it has been one of the
motivations for the work we report here. A
fuller recent discussion including a review
of the literature may be found in 
\cite{discreteness3_mjbm}.

As in our discussion of our finite system simulations 
above, it is necessary evidently to determine first what
extrapolation of the parameters in the problem 
corresponds to the VP limit. In the set of initial
conditions we have discussed, we have, as we discussed,
only a single parameter, the normalized shuffling $\delta$.
It is straightforward to see that variation of this
parameter does not give convergence to the VP limit
(just as variation of the single parameter $N$ in
the finite case did not define such convergence): 
at any value of $\delta$ an analysis of the dynamics
(see in particular \cite{sl3}) shows that 
there is always a phase of the evolution at
early times in which the forces on particles
are determined predominantly by their nearest
neighbors, which means that the mean-field 
VP limit is certainly not valid. Further for
small values of $\delta$ one can show, using
a perturbative treatment of the dynamics, that
there are measurable deviations from the
fluid linear theory limit at any wave-number
$k$ \cite{marcos_06}.

While the VP limit in the case of a finite system is an extrapolation
of particle number $N$, in the infinite system it cannot be defined in
this way (as $N$ is already infinite!). Instead it must be clearly be
defined as an extrapolation of the {\it particle density}, keeping the
quantities fixed which are relevant to the dynamics observed.  To make
sense of such a limit in the system we are studying it is evident that
we need some other length scale (with which to compare the mean
inter-particle distance). What is this length scale? Just as in the
finite case it is the dynamics itself which must define such a scale
if the VP limit is to be defined, as envisaged in the derivation of
this limit by coarse-graining \cite{buchert_dominguez}.  In the
infinite system we have studied it is clear, like in the finite
system, that the evolution can be stable only if the initial
fluctuations are kept fixed. All the density fluctuations in the
system cannot, however, be kept fixed when we vary particle density
(as shown, in particular, by the large $k$ behavior of the power
spectrum which is determined uniquely by this density).  Fluctuations
can nevertheless be kept fixed over some range of scales. Indeed this
is illustrated by the expression for the power spectrum of the SL
given above, Eq.~(\ref{eq:pk}), which shows that it suffices to vary
$\delta$ appropriately when $\ell$ changes in order to keep the long
wavelength fluctuations fixed.  Thus the length scale we assume to
exist, in order to define the VP limit, is, in complete analogy to the
finite case: we assume that the dynamics, in the spatial and temporal
range studied, are insensitive to fluctuations below this length
scale.  Such an extrapolation can be defined just as was described
above for the finite case, by breaking each of the particles in the
original lattice into a ``cloud'' of points which are redistributed
randomly on some scale $r_s$, which is naturally chosen to be of order
the inter-particle distance $\ell$ in the original distribution (but
can be determined a posteriori as was seen above). Alternatively it
can be defined in this case by an extrapolation in which only the
lattice spacing is decreased, varying $\delta$ to keep the small $k$
power spectrum the same, but with a fixed a cut-off scale $k=k_c$
above which power is filtered. More simply one can have the scale
$\varepsilon$ play the role of such a filter, as fluctuations below
this smoothing scale in the force are damped dynamically.  This is the
prescription which is most practically useful in cosmology, specifying
that the closeness to the (desired) VP limit should be tested for by
extrapolating $\ell \rightarrow 0$ at fixed $\varepsilon$.  Indeed on
this basis we expect to approximate well the VP limit with an $N$-body
simulation only when we work in the regime $\ell < \varepsilon$.  This
is {\it not} the current practice in cosmological simulations (see
e.g. Ref.~\cite{springel_05}).  In \cite{discreteness3_mjbm} we have
examined the issue of the resultant errors, placing non-trivial lower
bounds on them which show that they are certainly at a level relevant
to the precision required in coming years of these simulations. In our
study of the SL in \cite{sl3} we have shown that the asymptotic form
of the correlation function in this case is very similar to that which
emerges at early times when the evolution is clearly far from the VP
limit.  This suggests that careful further study of the relation
between the two regimes, $\ell < \varepsilon$ and $\ell >
\varepsilon$, even if numerically costly, will be necessary to resolve
these issues, which are of both theoretical and practical importance
in the theory of structure formation in the universe.

\section{Summary and conclusions}

We have described the phenomenology of the evolution of 
self-gravitating systems of particles from simple
one parameter classes of quasi-uniform initial 
conditions, as well as some basic theoretical results
which explain aspects of their behavior. In both cases
it can be said that very robust behaviors of the
``final states'' are observed --- the characteristic
profiles of the QSS in the finite system, and the
form of the asymptotic non-linear correlation 
function in the infinite system ---  but that in
both cases these behaviors are not at all 
understood. These are open theoretical problems
which it might be profitable to address also in
the context of even simpler toy models, e.g.,
the ``sheet'' model in one dimension (see
e.g. \cite{miller_1dreview} for a review of finite
systems, and 
\cite{miller+leguirriec, gabrielli+joyce+sicard_2008} 
for recent studies of the infinite system case). 

The particular class of initial condition we chose
for the finite case allowed us to explain the
meaning of the infinite space limit: the dynamics
of cosmological simulations of structure formation
in an infinite expanding universe as studied 
is simply the dynamics of clustering observed 
inside such a finite spherical system, initially
almost homogeneously expanding, in the limit
that the size of the sphere goes to infinity.
The static universe limit, on the other hand,
is defined most clearly as the infinite volume
limit of a finite system in which there is
a screened gravitational interaction, the
screening being very large itself in comparison 
to the scales up to which the system has
clustered. 

In both cases we have discussed the validity of
a description of the observed dynamics by the
VP limit, and the question of how to actually
test for such validity. This is of importance
both theoretically, as it is essential to know 
whether the VP equations provide the right 
framework in which to try to understand the
observed dynamics, and practically, as the 
goal of simulations of these kinds in cosmology
and astrophysics is usually to approximate 
the VP limit. In both systems discussed we
have given well defined prescriptions for the 
extrapolation to this limit. We have 
found numerically in the finite case that
such an extrapolation does indeed appear
to give stable results for the observed 
(macroscopic) dynamics, while in the
infinite case further numerical study is
required. We have emphasized that these prescriptions
require the introduction of a length scale which
is kept fixed as the particle density is
increased. We have identified this length
scale as the maximal scale down to which 
we need to keep initial density fluctuations
fixed in order to obtain the observed  
dynamics. While such a definition of the VP 
limit may be justified theoretically by
existing derivations (see e.g. \cite{buchert_dominguez}), 
it would be desirable that they made more rigorous 
in future work, in particular for the case of 
the infinite system limit.

We acknowledge the E. Fermi Center, Rome, for use of its computational
resources, and  Thierry Baertschiger and Andrea Gabrielli for 
collaboration on the results discussed in the second part of 
these proceedings. MJ also thanks Francois Sicard for useful 
conversations.

\section*{References}







\end{document}